\documentclass[preprint,journal]{vgtc}            

\usepackage[normalem]{ulem}
\usepackage{wrapfig}

\newcommand{\system}{\texttt{EVM}}
\newcommand{\circA}{\raisebox{.5pt}{\textcircled{\raisebox{-.6pt} {A}}}}
\newcommand{\circB}{\raisebox{.5pt}{\textcircled{\raisebox{-.8pt} {B}}}}
\newcommand{\circC}{\raisebox{.5pt}{\textcircled{\raisebox{-1pt} {C}}}}
\newcommand{\circD}{\raisebox{.5pt}{\textcircled{\raisebox{-1pt} {D}}}}

\newif\iftrack
\trackfalse 
\newcommand{\revision}[1]{\iftrack{\textcolor{red}{#1}}\else#1\fi}

\onlineid{1787}

\vgtccategory{Research}

\vgtcpapertype{system}

\title{\system: Incorporating Model Checking into \\ Exploratory Visual Analysis}

\author{Alex Kale, Ziyang Guo, Xiao Li Qiao, Jeffrey Heer, Jessica Hullman}

\authorfooter{
  \item
  	Alex Kale is with the University of Chicago.
  	E-mail: kalea@uchicago.edu.
   \item
  	Ziyang Guo is with Northwestern University.
  	E-mail: ziyangguo2027@u.northwestern.edu.
   \item
  	Xiao Li Qiao is with Northwestern University.
  	E-mail: emqiao@gmail.com.
   \item
  	Jeffrey Heer is with the University of Washington.
  	E-mail: jheer@uw.edu.
   \item
  	Jessica Hullman is with the Northwestern University.
  	E-mail: jhullman@northwestern.edu.
}

\abstract{
    Visual analytics (VA) tools support data exploration by helping analysts quickly and iteratively generate views of data which reveal interesting patterns.
    However, these tools seldom enable explicit checks of the resulting interpretations of data---e.g., whether patterns can be accounted for by a model that 
    \revision{implies a particular structure in}
    the relationships between variables.
    We present \system{}, a data exploration tool that enables users to express and check provisional interpretations of data in the form of statistical models.
    \system{} integrates support for visualization-based model checks by rendering distributions of model predictions alongside user-generated views of data.
    In a user study with data scientists practicing in the private and public sector, we evaluate how model checks influence analysts' thinking during data exploration.
    Our analysis characterizes how participants use model checks to scrutinize expectations about data generating process and surfaces further opportunities to scaffold model exploration in VA tools.
}


\teaser{
  \centering
  \includegraphics[width=\linewidth]{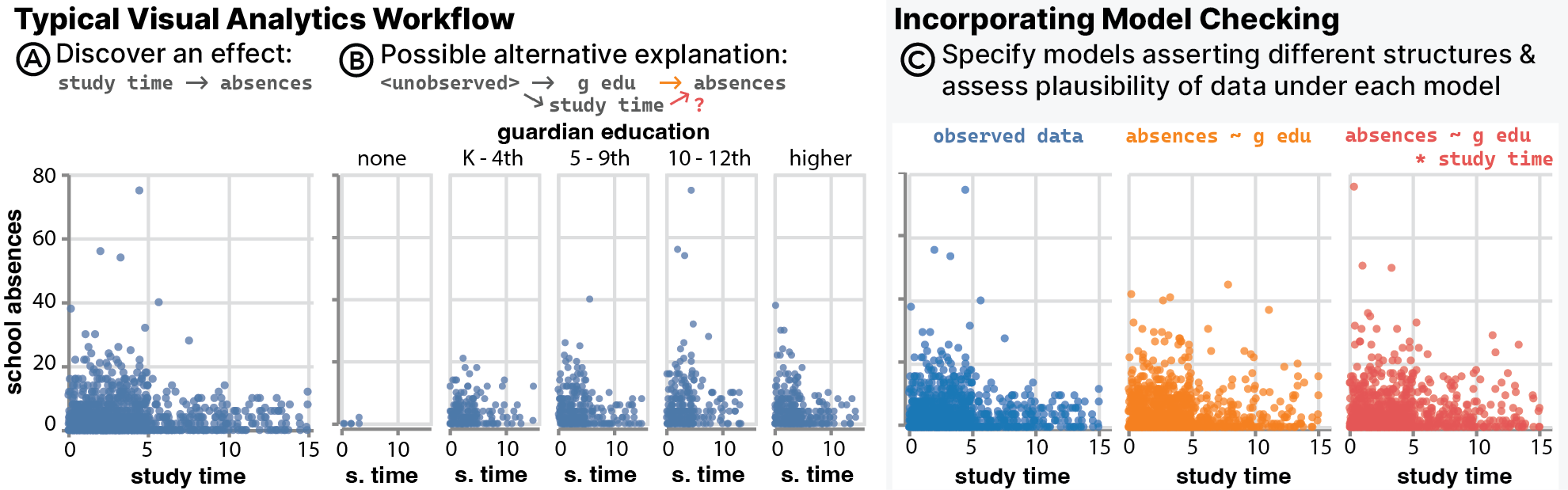}
  \caption{
  Model checks modify the typical visual analytics workflow by enabling users to assess the plausibility of 
  \revision{interpretations of}
  discovered patterns. \circA{} The analyst discovers that accounting for time spent studying appears to help explain student absences. \circB{} The analyst facets this view by the highest level of education achieved by each student's guardian. He wonders if study time is predictive of absences after accounting for guardian education. \circC{} The analyst specifies models 
  \revision{asserting}
  that absences are explained by either: guardian education alone (orange); or guardian education and study time (red). Seeing that predictions from the second model do a better job of capturing the largest numbers of absences, he concludes that both guardian education and study time are important explanatory variables.
  }
  \label{fig:teaser}
}

\graphicspath{{figs/}{figures/}{pictures/}{images/}{./}} 

\usepackage{mathptmx}                  

\begin{document}


\firstsection{Introduction}
\label{sec:intro} 

\maketitle

Data analysts use exploratory visual analysis (EVA) tools such as Tableau to check their 
\revision{understanding of}
data, discover patterns,
and seek potential explanations for those patterns.
For example, imagine an analyst, Juan, contracted to investigate what factors contribute to school absences in a local school district.
During exploration, Juan discovers an interesting looking pattern (Fig.~\ref{fig:teaser}~\circA) where absences are associated with the number of hours students spend studying.
Juan wonders what could explain this pattern.
He begins faceting by other variables in the dataset provided by the school and notices that the highest level of education achieved by each student's guardian helps to explain large numbers of absences among students who study less (Fig.~\ref{fig:teaser}~\circB).
If Juan wants to scrutinize 
\revision{his interpretation of this pattern}
further, and he is comfortable with statistical programming, he might switch to a programming language like R or Python to specify and compare the predictions of models that do and do not 
\revision{assert this relationship.}
This kind of provisional ``model checking''~\cite{hullman2021hdsr} 
\revision{could help Juan rule out some interpretations as less likely, or explore how variations on his interpretation might better capture what he sees.}

Given their broad appeal to non-programmers, many users of visual analysis (VA) tools may not be comfortable shifting to statistical tools to do preliminary model development and checking. 
This points to a gap in current design approaches for EVA tools: they offer little support for helping analysts 
\revision{reason more explicitly about their provisional interpretations.}
Recent work~\cite{hullman2021hdsr} critiques the standard approach to designing visual analysis tools as implying a misconception that data exploration is ``model free'', in the sense that we rely primarily on visualizations for surfacing patterns and hypotheses early in analysis and rely primarily on modeling for confirmatory statistical inference late in analysis.
A consequence of this ``model free'' account of data exploration is siloed tools for exploratory and confirmatory analysis, leaving EVA users under-equipped to resolve ambiguity about the underlying data generating process.

Imagine instead that an EVA interface enabled Juan to (1) fit a series of  
\revision{models asserting various plausible assumptions and structures about the data}
and (2) visualize their predictions in comparison to 
\revision{observed}
data (Fig.~\ref{fig:teaser}~\circC).
Explicitly comparing models encoding different 
\revision{data interpretations}
can afford better-calibrated confidence in claims that Juan might make about the data generating process.
Without such an explicit representation of expectations, analysts like Juan must \textit{imagine what it could look like if}, e.g., only guardian education influences student absences and whether the observed data are plausible under that model, imagined counterfactuals that some 
recent work (e.g.,~\cite{Kale2022causal}) suggests are difficult for many to accurately incorporate into visual inferences.
At the same time, once an EVA tool makes it easy to generate model predictions for comparison with data, it is possible that the added formality provides a sense of false assurance to analysts, exacerbating risks of
\revision{post-hoc inferences or overfit explanatory models, similar to p-hacking~\cite{Simmons2011} or model selection via R-squared~\cite{yarkoni2017}.}

We explore the promise and pitfalls of explicit modeling support integrated in a visual data exploration tool via  
Exploratory Visual Modeling (\system{}), an open-source prototype EVA application. \system{} enables users to quickly specify statistical models representing plausible explanations of discovered patterns in data, and compare predictions from these models to the observed data. 
The key role of visualization in \system{} is \textit{showing discrepancies between patterns in data and expectations}, a design pattern 
\revision{dubbed \textbf{``model checks''} by Hullman and Gelman~\cite{hullman2021hdsr}.}

\system{} contributes a drag-and-drop interface for chart construction (c.f., Polaris~\cite{stolte2002polaris}) with model checking features including: an interactive ``model bar'' that enables analysts to flexibly specify regression models, a back-end layer that fits user-specified models and samples appropriate predictive distributions to show in the browser, and 
\revision{carefully designed}
defaults for model check visualizations that juxtapose user-generated views of data with model predictions.
This interface enables users to quickly specify and check a wide variety of regression models against data on-the-fly, just as conventional EVA tools enable quick chart construction. 
Additionally, we present an evaluation of how 12 data scientists practicing in the private and public sectors use model checks when they are incorporated into EVA workflows.
We characterize how model checks change participants' exploratory data analysis behavior compared to a baseline condition where they use a simple VA tool without model checking functionality, finding that model checks evoke different analysis behaviors depending on the user's previous 
\revision{experience with}
modeling.
We discuss new design requirements for model checks in EVA, highlighting cognitive pitfalls of model-based data exploration, possible roles for model recommendation, and opportunities around new programming tools for visual modeling.
\section{Related work}
\label{sec:related}
We present relevant literature on graphical statistical inference and interactive model selection to contextualize our contributions.

\subsection{Graphical statistical inference}
\label{sec:related-gsi}
Graphical statistical inference refers to visual methods for judging how well a statistical model describes observed patterns in data, which we refer to as \textit{``model checks''} following Hullman and Gelman~\cite{hullman2021hdsr}.
Tukey motivated visualizing model residuals to inspect where a provisional model might be wrong during exploratory data analysis~\cite{tukey1966data}.
Similarly, common model diagnostic tools such as QQ-plots~\cite{Loy2016} create visual tests that a model's assumptions are satisfied. 

The best-known visualization formulation of model checks may be the visualization lineup protocol~\cite{buja1987data,Buja2009,Wickham2010}, in which many plots of 
\revision{simulated}
data are generated from a ``null model'' (i.e., representing a null hypothesis) and an impartial observer is asked to pick out a plot of the real data among the set of 
\revision{``null plots.''}
The lineup procedure has been analogized to a formal statistical test, and shown to have equivalent power or even better power than a conventional statistical test in some scenarios~\cite{majumder2013validation}. 
However, the strict analogy between the lineup and a statistical test can be hard to ensure because 
\revision{creating null plots is a non-trivial challenge~\cite{VanderPlas2017,Vanderplas2021}, and the need for impartial observers is impractical.}
Others have proposed alternative analogies for graphical inference, such as Bayesian cognition, as a means of guiding visualization design and research~\cite{hullman2021hdsr,Kim2019,kim2020bayesian,wu2017towards}.
Inspired by Bayesian workflows, where visualizations are the primary means by which models are interrogated~\cite{gabry2019,gelman2020bayesian}, Hullman and Gelman~\cite{hullman2021hdsr} discuss how 
\revision{model check visualizations}
can have value for helping analysts understand \textit{for which cases a model is wrong about the data}, without evoking an analogy to a statistical test or attempting to provide guarantees about error rates.
In this approach, rejecting a model is not considered a win so much as realizing from a visualized model check what features of observed data remain yet to be explained.

While most interpretations of visualizations might be likened to checking an implicit model representing the viewer's expectations about the data~\cite{gelman2003,gelman2004exploratory}, attempts to design graphical user interfaces for visual analysis (`VA tools' hereafter) that make it easy to check provisional models have been much less prevalent than design approaches that emphasize interaction with observed data~\cite{hullman2021hdsr}. 
For example, while Tableau Software supports very simple regression modeling and construction of uncertainty intervals, at the time of this writing plotting residuals requires multiple data transformation and visualization steps. 
In part because of this lack of integration of model checking with VA tools, 
\revision{analysts may not always scrutinize}
patterns discovered using such tools 
\revision{to ask how exactly they might arise.}
Exceptions include 
\revision{early research systems developed by and for statisticians~\cite{becker1987dynamic,velleman2012datadesk}, NorthStar created for predictive modeling~\cite{Kraska2018}, and recent research systems developed to study novel interfaces for eliciting analysts' expectations via natural language~\cite{Choi2019b} and sketching~\cite{Koonchanok2021}.}
\revision{In contrast to these efforts, we developed \system{} to study how model check visualizations might benefit the broader populations of analysts that tools like Tableau target, assuming neither that our users would be statisticians nor that 
realizing model checking necessarily requires a new elicitation medium.
} 

\subsection{Visually-aided model selection}
Descriptive accounts of exploratory visual analysis (EVA, e.g.,~\cite{battle2019-characterizing-eva}) acknowledge that it often alternates between open-ended tasks (e.g., flipping through filters looking for something interesting to explore a space of theories or models, a.k.a, abduction proper~\cite{devezer2020case,oberauer2019addressing}) and more focused exploration (e.g., trying to formulate and validate a hypothesis). 
Recent work in computer science~\cite{pu2018garden,Zgraggen2018,zhao2017controlling} analogizing EVA to a multiple comparisons problem emphasizes what Tukey~\cite{tukey1972data} referred to as ``rough confirmatory analysis.'' 
In this stage, visual analysis plays a classification role in helping an analyst distinguish between signals that are so apparent that statistical modeling is not needed, versus where noise and confounding are so great that confirming perceived patterns is hopeless. \system{} is designed specifically to support this rough confirmatory stage, in which analysts \revision{rely on} their eyes to make often difficult judgments about signal versus noise ratios. 

Tukey stressed multiplicity as a key issue in this intermediate stage of analysis, which proceeds a stage of initial exploration in which probability is not of interest, and precedes confirmatory analysis. 
For example, an analyst should be wary of ``How many things might have been looked at? How many had a real chance to be looked at? How should the multiplicity decided upon, in answer to these questions, affect the resulting confidence sets and significance levels?''~\cite{tukey1972exploratory}. 
\revision{Whereas previous research~\cite{Choi2019a,Choi2019b,Koonchanok2021} attempted to avoid risks of post-hoc inference by forcing analysts to specify models before seeing the results of queries, and suggested further mitigations through automated adjustment of test statistics, we opted to focus \system{} on making support for model check visualizations as seamless as possible, without even providing numerical model summaries like p-values for users to exploit.}
\revision{We built \system{} to investigate how VA users rely on unconstrained visual checking to search a space of plausible models, rather than presupposing a hypothesis testing framework where elicited models necessarily reflect a user's best-guess expectation.}
The multiple comparisons problem has led to valuable suggestions for EVA systems like holdout sets~\cite{Zgraggen2018}, which would be natural to support in future iterations of \system{}.


Other research prototypes~\cite{guo2023causalvis,Wang2016,Wang2018,Xie2020} have been developed to support causal inference by representing user-defined queries in terms of directed acyclic graphs.
These tools integrate data mining approaches into VA tools with the intention of helping users explore the plausibility that various causal structures explain their data.
Although \system{} has similar goals insofar as we aim to promote scrutiny about interpretations of data, we avoid automated modeling approaches in \system{} based on the design philosophy that visual model checking will be most meaningful when models originate from users' expectations. 

\section{Design requirements}
\label{sec:design}
We designed and implemented \system{} to support model checking during exploratory visual analysis.
We describe the design rationale for \system, highlighting where we envision model checks adding value to visual data exploration workflows.

\noindent
\textbf{Promote generative thinking.}
Analysts derive meaning from patterns they discover during visual data exploration based on how these patterns match or contradict their expectations.
To evaluate possible patterns and expectations, analysts must answer the question, \textit{``What would a new data sample look like given my provisional
\revision{model}?''}
However, visual data exploration tools usually do not support generation of new hypothetical samples, leaving analysts to imagine how these might look.
A primary design hypothesis behind EVM is that integrating model checking into visual data exploration will promote more explicit consideration of how data might have been generated. 

\noindent
\textbf{Pattern-seeking, not data mining.} 
Visual data exploration tools lend themselves to false discoveries~\cite{Zgraggen2018} in part because they make it easy to view data but hard to check 
\revision{interpretations}
or connect visual patterns to expectations.
Analysts may be distracted or mislead by small visual details on charts unless they seek these details deliberately based on questions about their data. 
For this reason, \system{} avoids serving up comparisons or views of data that the user has not requested.

\noindent
\revision{\textbf{Eliciting regression models.}}
In order to check users' 
\revision{provisional data interpretations,}
 we require a computational representation  
\revision{of the assumptions and structure implied by an explanation.}
Part of the design philosophy behind 
\revision{model check visualizations is that regression models can provide a shared abstraction for humans and machines, and that models need not represent either a user's best-guess of the data generating process or a hypothesis test in order for visualized model predictions to provide a useful reference for making sense of data~\cite{hullman2021hdsr}.}
For this reason, \system{}  
\revision{enables users to specify regression models using}
R syntax for model formulae~\cite{Wilkinson1973,Pinheiro2020,gamlss2005} 
through a component called the model bar (see Section~\ref{sec:system-functionality}).
\revision{Given that}
such regression formulae are shown to be a successful ``interface'' via their frequent use in the sciences, 
\revision{we assume they will be suitable for users with a}
wide range of backgrounds and experiences with statistics.

\noindent
\textbf{Model expansion workflow.}
One perspective in statistical theory suggests that tools should promote a modeling workflow where analysts consider and check \textit{incremental} changes to models reflecting their provisional beliefs about data~\cite{gabry2019}. 
Although there is some contention that such forward stepwise selection of predictors could lead to biased parameter estimates arising from multiple comparisons and overfitting~\cite{miller1984,harrell2001}, this is a risk primarily when the goal is hypothesis testing. 
In contrast, \system{} is intended for visually scrutinizing 
\revision{data interpretations in order to 
explore
a space of plausible models, not for hypothesis testing.} 
While analysts' difficulty in clearly separating exploratory and confirmatory analysis can contribute to overfit explanations~\cite{gelman2013garden}, and may also be a risk using \system, a goal of \system{} is to integrate better tools for preliminary winnowing of bad 
\revision{interpretations of visual findings}
that might otherwise go unchecked.
To this end, a model expansion workflow helps analysts work up to complex models in terms of simpler models that 
\revision{assert}
a subset of the same 
\revision{structure.}
\system{} facilitates this sort of cumulative assessment of what is and isn't helpful 
\revision{to model.}

\noindent
\textbf{Decouple models from visualizations.} 
It can be tempting to draw analogies between statistical models and visualizations (e.g., that Fig.~\ref{fig:teaser}~\circB~necessarily implies a model assuming an interaction between study time and guardian education), due to representational similarities at the software abstraction level between models~\cite{Wilkinson1973,Pinheiro2020,gamlss2005} and visualizations~\cite{Wilkinson2005}. 
However, in developing \system{} we quickly realized that a more appropriate way to describe the relationship between models and visualizations is many-to-many: a model can imply multiple visual checks and a visualization can map to multiple model specifications\revision{~\cite{talbotCommunication}}.
Hence \system{} decouples model specification from chart specification, allowing visualizations 
\revision{to}
show variables that are not predictors in a model, and 
\revision{models to}
learn relationships which are not directly visualized. 
\revision{This enables use cases like Figure~\ref{fig:teaser}~\circC~where model check visualizations can help users reason about the higher-dimensional patterns behind problematically low-dimensional~\cite{Wickham2015} visualizations.}

\noindent
\textbf{Graft model outputs to user-specified visualizations.}
Model checking is inherently visual, but few principles exist for prescribing how to show model outputs alongside 
\revision{observed data~\cite{Wickham2015},}
including how to sample from the model output and how to facilitate visual comparisons.
We strove for smart defaults such that \system{} juxtaposes model predictions with any user-generated visualization by simply adding an adjacent subplot for each fitted model.
\system{} always facets subplots of the model predictions in a layout that preserves the ability to \textit{compare observed and predicted outcomes on a common scale}.
We show model predictions as animated hypothetical outcome plots or HOPs~\cite{Hullman2015} since this uncertainty visualization technique is helpful for showing reference distributions~\cite{Kale2019hops} and can be applied to any user-generated chart.
When the analyst chooses to check a model that doesn't make sense (e.g., using a model that assumes discrete outcomes on continuous data), the resulting model check visualization sometimes becomes ill formed, signaling to the analyst that something has gone wrong.
After discovering these failure modes during informal testing, we decided not to prevent them based on their value for detecting 
\revision{flawed models.}
\section{\system: Exploratory Visual Modeling}
\label{sec:system}
We implemented \system{} as a single-page web application, where users can generate views of data and models to check, connected to an R server that fits models and extracts predictions from them. 
The reader can interact with the prototype at \href{https://mucollective.github.io/evm/}{https://mucollective.github.io/evm/} and find the project repository at \href{https://github.com/MUCollective/evm/}{https://github.com/MUCollective/evm/}.

\begin{figure}[t]
    \centering
    \includegraphics[width=3in]{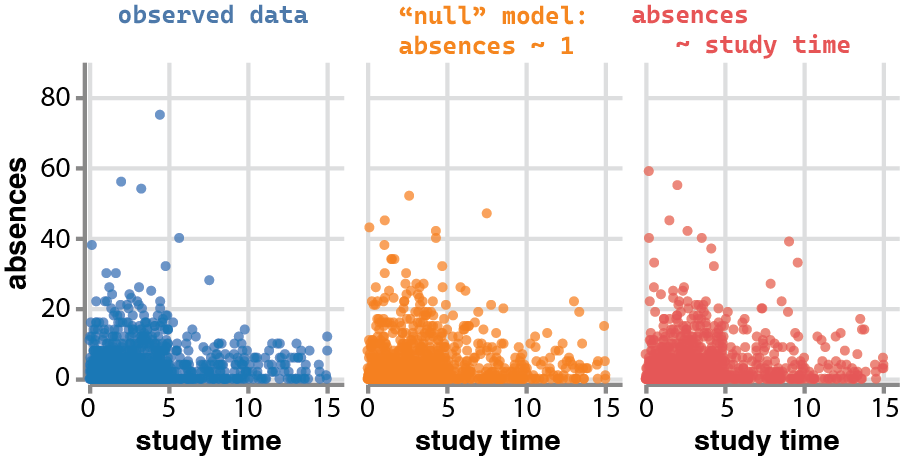}
    \setlength{\abovecaptionskip}{3pt}
    \setlength{\belowcaptionskip}{-14pt}
    \caption{Model check showing what the data would look like if \texttt{absences} was not associated with \texttt{study\_time} (orange) vs if it was (red). Predictions from the two models look very similar.
    }
    \label{fig:null-mc}
\end{figure}

\begin{figure*}
    \centering
    \includegraphics[width=6.9in]{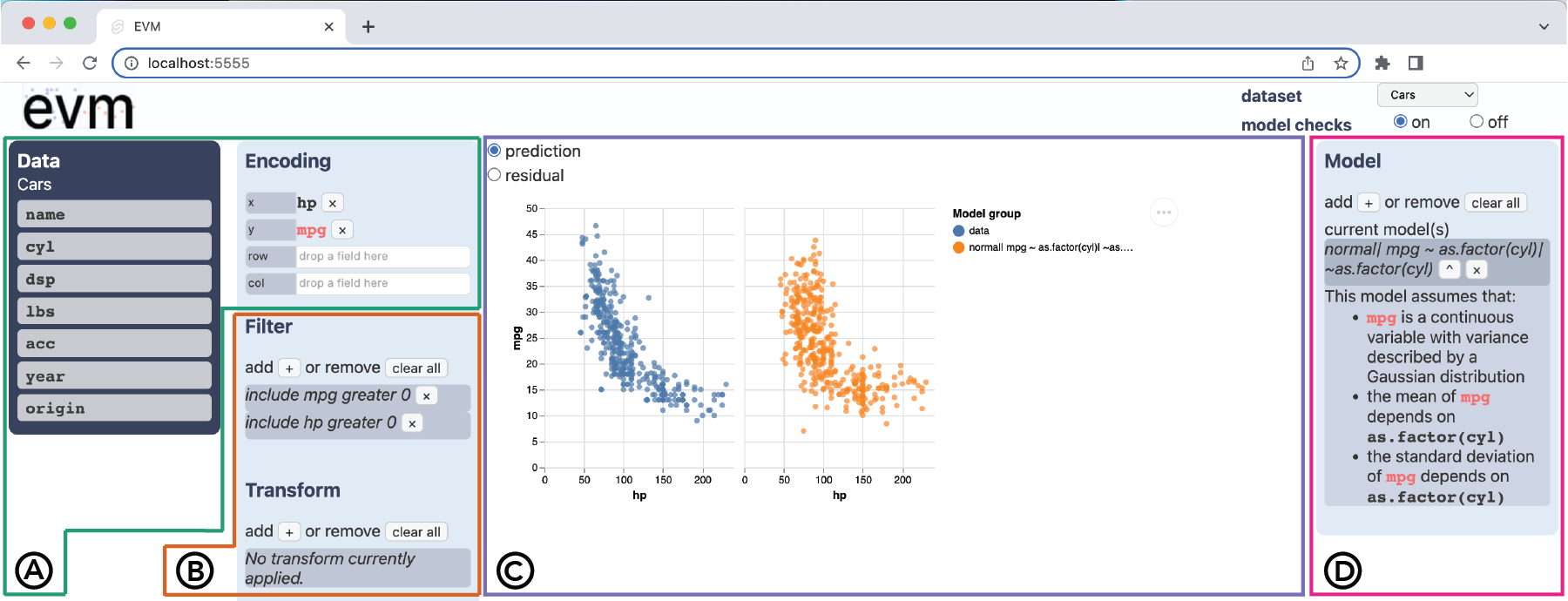}
    \setlength{\abovecaptionskip}{3pt}
    \setlength{\belowcaptionskip}{-14pt}
    \caption{A screenshot of \system{}, annotated to show different components described in Section~\ref{sec:system-functionality}:
    \circA{} Shelf construction; 
    \circB{} Filters \& transformations; 
    \circC{} Chart panel; and
    \circD{} Model bar.
    This model check assumes that the number of miles per gallon a car gets \texttt{mpg} is normally distributed with mean and variance both influenced by the number of cylinders in a car's engine \texttt{cyl}.
    Plotting horsepower \texttt{hp} against \texttt{mpg} results in a model check of how well the trade-off between \texttt{hp} and \texttt{mpg} is explained by the variable \texttt{cyl}, which is not shown directly in this view.
    }
    \label{fig:evm-expo}
\end{figure*}

\subsection{Usage scenario}
\label{sec:system-scenario}
Consider how Juan might interact with \system{} to investigate factors that affect student absences (see Section~1). 
Initially, Juan might use \system's drag-and-drop interface (Fig.~\ref{fig:evm-expo}~\circA) to construct a series of bar charts, strip plots, and scatterplots examining each predictor variable in the dataset and its relationship with the outcome of interest, \texttt{absences}.
This initial tour of the data would reveal interesting potential relationships such as associations between student \texttt{absences} and variables like weekly hours of \texttt{study\_time} or level of guardian education (\texttt{g\_edu}).

In order to scrutinize these relationships, Juan uses \system's model bar (Fig.~\ref{fig:evm-expo}~\circD) to express provisional interpretations to check against the data.
First, Juan \textit{evaluates the plausibility of the 
\revision{interpretation}
} that \texttt{study\_time} helps explain \texttt{absences}. 
To do this, he specifies two models to check against the data: one ``null'' model 
\revision{asserts}
no relationship between \texttt{study\_time} and \texttt{absences}; the other model
\revision{asserts}
that \texttt{study\_time} is predictive of \texttt{absences} (Fig.~\ref{fig:null-mc}).
In both models, Juan assumes that the empirical distribution of \texttt{absences} can be approximated by a negative binomial distribution because absences are an overdispersed count outcome.
By comparing the data distribution to predictions from both of these models in a model check, Juan becomes less convinced that \texttt{study\_time} is an important predictor.

Juan 
\revision{wonders if}
\texttt{study\_time} might only seem predictive because 
\revision{of correlation}
with 
other explanatory variables.
Returning to a 
\revision{visual exploration}
workflow, he starts faceting the relationship between \texttt{absences} and \texttt{study\_time} by other factors. 
Juan discovers that the pattern looks stronger, with 
higher numbers of absences overall, when students have guardians with higher levels of education (Fig.~\ref{fig:teaser}~\circB).
Perhaps both \texttt{study\_time} and \texttt{g\_edu} reflect some unobserved factor such as a family's socioeconomic status.
Juan wonders if it is still important to consider \texttt{study\_time} after accounting for \texttt{g\_edu}.

To better \textit{understand correlated predictors}, Juan sets up a series of models to check against the data. 
He starts with a model 
\revision{asserting}
that only \texttt{g\_edu} influences \texttt{absences}.
\system{} juxtaposes predictions from this model against Juan's scatterplot of the relationship between \texttt{study\_time} and \texttt{absences} (\revision{Fig.~\ref{fig:teaser}~\circC}, left vs middle), revealing that the pattern in the data can be roughly accounted for by a model that 
\revision{asserts}
only an effect of \texttt{g\_edu}.
Juan tries adding another model 
\revision{asserting}
that both \texttt{g\_edu} and \texttt{study\_time} are predictive of \texttt{absences} to investigate whether using \texttt{study\_time} as a predictor improves the model fit at all.
The resulting model check shows that a model 
\revision{asserting}
influences of both \texttt{g\_edu} and \texttt{study\_time} does a better job of predicting 
\revision{the case with the largest number of absences}
(\revision{Fig.~\ref{fig:teaser}~\circC}, left vs middle vs right).
\revision{Juan's interpretation of this model check depends on whether he thinks of this case as an outlier or the tail of the absences distribution.} 
\system's model checks help Juan arrive at the conclusion that, although \texttt{g\_edu} and \texttt{study\_time} are correlated, these predictors likely contain some non-overlapping information, something that he could not ascertain from exploratory visual analysis alone.

\subsection{Overview of functionality}
\label{sec:system-functionality}
The basic visual analytics functionality of \system{} resembles that of systems like Tableau Software. We based this aspect of \system's design on PoleStar~\cite{voyager2017}, a research prototype developed to mimic the interaction model of Tableau for the purpose of user testing.
To generate visualizations in \system, users rely on a \textbf{shelf construction} interface (Fig.~\ref{fig:evm-expo}~\circA) where they drag ``pills'' representing variables in a dataset onto ``shelves'' representing x, y, row, and column encodings (in the sense of the grammar of graphics~\cite{Wilkinson2005}).
The resulting visualizations appear in the \textbf{chart panel} at the center the display (Fig.~\ref{fig:evm-expo}~\circC). 
Smart defaults determine the chart types that render in the chart panel depending on the data types of the variables on the x and y encodings:
\begin{itemize}[noitemsep]
    \item \textit{Bar charts} show univariate distributions for discrete variables.
    \item \textit{Strip plots} show univariate distributions for continuous variables, and bivariate distributions for continuous vs. discrete variables.
    \item \textit{Scatterplots} show bivariate distributions for continuous variables.
    \item \textit{Heatmaps} show bivariate distributions for discrete variables.
\end{itemize}
\system{} facets any of these chart types into a \textit{trellis plot}~\cite{Tukey1977,Becker1996}, consisting of multiple subplots arranged along the vertical and/or horizontal span of the chart panel, when the user defines a row and/or column encoding.
In addition to chart specification, users of \system{} can add \textbf{filters and transformations} by clicking the \texttt{+} icons (Fig.~\ref{fig:evm-expo}~\circB).
When adding filters, the user selects a variable based on which they will either include or exclude values less than (or equal to), greater than (or equal to), or (not) equal to a criterion.
When adding transforms, the user selects a variable to which they can apply either a log odds or log transformation, two options provided in \system{} because they are used to handle bounded distributions in logit normal and log normal models, respectively (see below). 
One can remove filters and transforms by clicking the \texttt{X} icons next to each filter or transform or by clicking \texttt{remove all}. 
For simplicity, \system{} always applies filters before transforms and applies both in the order they are specified.

\begin{figure}
    \centering
    \includegraphics[width=3in]{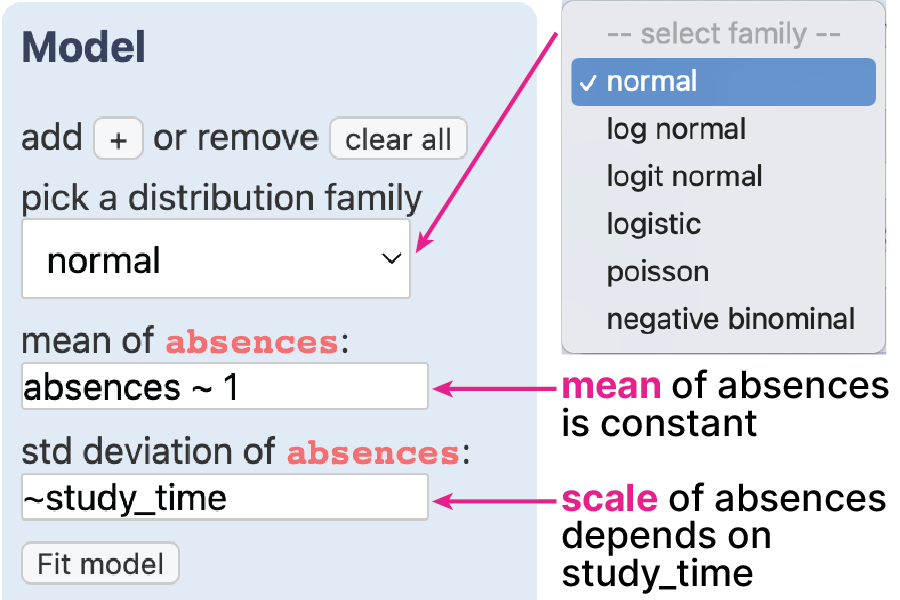}
    \setlength{\abovecaptionskip}{3pt}
    \setlength{\belowcaptionskip}{-14pt}
    \caption{The model bar specifying a model check where absences are assumed to be normally distributed with a constant mean and variance dependent on \texttt{study\_time}.}
    \label{fig:model-bar}
\end{figure}

The primary innovation of \system{} is to add model-checking functionality to this style of visual analytics system.
Users can specify models to check against visualizations using the \textbf{model bar}, an interface element not found in prior exploratory visualization tools (Fig.~\ref{fig:evm-expo}~\circD).
The model bar employs a similar design pattern to the filter and transform interfaces, in that users can add or remove models from the current set of candidate models.
When the user chooses to add a model to the model bar, they select a \textit{distribution family} from the following options: \textit{normal (a.k.a. Gaussian)} for unbounded continuous outcomes; \textit{log normal} for right-skewed continuous outcomes with a lower bound at zero; \textit{logit normal} for continuous outcomes bounded at zero and one; \textit{logistic} for binary outcomes; \textit{Poisson} for count outcomes; and \textit{negative binomial} for overdispersed count outcomes. 
We selected these families to cover common distributions of outcome variables without redundancy (see Appendix in Supplemental Materials).
Based on the model family chosen by the user, \system{} elicits a \textit{model specification} in Wilkinson-Pinheiro-Bates syntax~\cite{Wilkinson1973,Pinheiro2020}, providing separate text boxes for location and scale sub-models (Fig.~\ref{fig:model-bar}) when the user selects a distribution family with an explicit scale parameter. 
\system{} parses these model specifications into 
\revision{assumptions and asserted structures and displays these as a list of natural language descriptions in the model bar.}


\subsection{Implementation}
\label{sec:system-implementation}
We implemented \system{} as a Svelte application that runs in the browser and connects to an OpenCPU~\cite{OpenCPU} deployment, which fits and processes models in R.
We use Vega~\cite{vega2016} to generate visualizations in \system{}.
We also wrote a custom R package named \texttt{modelcheck}, which
contains functions to run a specified model in the selected family, add model predictions to the input data frame, calculate residuals, and merge together outputs from other operations without duplicating entries. 
We rely on a combination of base R and \texttt{tidyverse}~\cite{tidyverse} for data wrangling, as well as \texttt{gamlss}~\cite{gamlss2005} models which fit quickly enough to keep latency suitably low for an interactive system.

In addition to fitting models, \texttt{modelcheck} handles two critical steps for uncertainty propagation that can pose challenges in modeling workflows: (1) sampling learned parameter values from the model to construct a predictive distribution and (2) back-transforming these predictions into the units of the data that users pass into the model bar.
This guarantees that model check visualizations compare data and model outputs on the same scale.
For example, predictions from log and logit normal model families are back-transformed using exponential and logistic inverse-link functions, respectively.
This implementation facilitates quick model iteration in \system's user interface; typically users would need to write analysis scripts to achieve similar model checks. 

\section{User study}
\label{sec:eval}
We designed \system{} for visual analytics (VA) practitioners whose backgrounds in statistics 
vary, from having taken an introductory statistics course at one time to fluidity with statistical tools.
In order to assess whether incorporating model checks into an interface like \system{} would benefit at least some people currently using VA tools, we ran a user study targeting practicing data workers who were familiar with statistical models but who,
\revision{unlike statisticians,}
typically might not incorporate modeling into exploratory workflows. 
Given the exploratory nature of our study, we conducted 
think-aloud\revision{s}
followed by open-ended conversational interviews with users, rather than attempting a controlled experiment 
\revision{targeting anticipated benefits of model checking.}
To investigate 
\revision{how model checking changes VA workflows,}
we characterized users' analysis behaviors at baseline using a simple VA tool without model checking functionality as well as using model checks in \system{}.

\subsection{Participants}
\label{sec:eval-participants}
We recruited practicing data analysts ($n=12$) to use \system{}, drawing primarily from our professional network via Twitter and email. 
To be eligible, participants needed to (1) work with data regularly, (2) be familiar with visual analytics tools like Tableau, and (3) have some previous experience using regression models.
We reasoned that recruiting real data analysts would provide greater ecological validity to any conclusions we draw. 
Of our 12 participants, 2 were academic researchers in computer science, 5 were data scientists working on business intelligence, 2 were data analysts working in healthcare, and 3 performed data-intensive work in government agencies or non-profits.
\revision{This sample emphasized practices of non-statisticians using VA tools.}

\subsection{Datasets}
\label{sec:eval-datasets}
We provided each participant with two cleaned datasets to explore using \system{}, without and with model checks enabled. 
Our study design required that these datasets were realistic without requiring specialized domain knowledge to explore, contained non-trivial structure for participants to discover, and were roughly equal in size.
We used two real datasets on forest fires~\cite{forest-fires-data,ml-datasets-uci} and student absences~\cite{absences-data,ml-datasets-uci} in Portugal, which met the first two requirements. 
To limit potential confounding effects of dataset size on analysis behavior, we matched the number of variables available for exploration by dropping selected variables until there were ten variables per dataset, and we matched the number of records in the datasets by dropping rows at random from the larger dataset (student absences) until both datasets contained 517 observations.
We also used aggregation and sampling procedures on a few variables to match the number of variables in each dataset that were discrete versus continuous (see Supplemental Materials).

\subsection{Interview protocol}
\label{sec:eval-protocol}
Our interview sessions were structured as a pre-post study design, where participants used a prototype visualization system with and without model check functionality, followed by a debriefing interview.
Each session spanned 90 minutes total, split into three 30-minute sections: \textit{think-aloud baseline}, \textit{think-aloud with model checks}, and \textit{debrief}.
\revision{We focus here on the procedure and goals of our evaluation; see Supplemental Materials for full details of the user study protocol.}

\noindent
\textbf{\textit{Think-aloud baseline}.}
In the first 5 minutes, we introduced participants to a version of \system{} where \textit{model checks were not enabled}. 
This entailed a demonstration of chart specification, data filtering, and transforms. 
In the next 25 minutes, participants explored one dataset in a think-aloud protocol using \system{} without model checks enabled. 
This step provided a baseline for characterizing data exploration behavior with a conventional VA tool at an individual level.

\noindent
\textbf{\textit{Think-aloud with model checks}.}
The interviewer enabled model checking and spent 5 minutes using the cars dataset~\cite{cars-data,ml-datasets-uci} to demonstrate potential model checking use cases investigating the plausibility of assumed relationships and correlations between predictors (see Section~\ref{sec:system-scenario}).
For example, to show how model checks can elucidate correlations between predictors, we showed the scenario in Figure~\ref{fig:evm-expo}~\circC.
This was followed by another 25 minutes of think-aloud data exploration on a second dataset, this time using \system{} \textit{with model checks enabled}. 
This second round of data exploration served as an intervention condition, assigned within-subjects to assess changes from baseline.

During think-aloud sessions, participants were instructed
\revision{to spend 25 minutes exploring one of two datasets looking for potential influences on either area burned in forest fires or student absences.}
\revision{We asked participants to tell us about any observations or patterns they felt were worth having a colleague follow up on.}
The interviewer 
\revision{spoke only to prompt participants}
to say what they were thinking, to answer direct questions, and occasionally to help participants get unstuck if they encountered a bug or confusing edge case.
Some participants, especially those who were less familiar with implementing regression models in R, needed clarification about model notation and underlying assumptions. 
The interviewer answered these questions.
When participants hesitated or got confused about model specification, the interviewer made a note and asked about these instances later in the interview.

\revision{In}
the two think-aloud\revision{s,} 
the pairing of 
datasets with 
interface conditions (i.e., model checks disabled vs. enabled) was counterbalanced across subjects, 
\revision{but}
the order of interface conditions was not. 
Our rationale was to control for artifacts of exploring a particular dataset while also avoiding a complex experimental design. 
Counterbalancing the order of interface conditions would have told us 
whether users seemed to explore data differently in a typical 
\revision{VA}
tool after exploring data with model checks---\revision{a learning effect that is not the focus of our evaluation.}

\noindent
\textbf{\textit{Debrief}}
The last 30 minutes of each interview involved a semi-structured conversational interview with the participant about \system{}.
The semi-structured interview followed an interview guide, which consisted of the following lines of questioning:
\begin{itemize}[noitemsep]
    \item \textbf{Utility of model checks.} In what ways (if any) did you use model checks to help you think about the dataset? What specific visual cues on the resulting chart (if any) were interesting or helpful?
    \item \textbf{Generative thinking.} Did you find yourself thinking about the data generating process, or the underlying relationships that might explain the patterns you saw in the data? What kinds of assumptions (if any) did you make about the dataset? Did using model checks make these assumptions more salient or concrete?
    \item \textbf{Expressiveness and usability.} Did you have any difficulty using the model bar to express and check provisional 
    \revision{interpretations of}
    data? What if anything made it hard to use? What if anything do you think would make this kind of functionality easier to use?
\end{itemize}
In each line of questioning, the interviewer referred back to specific examples of situations where the participant either created or attempted to specify a model check. 
The goal of this discussion was to elicit participants' reflections on how \system{} influenced their analysis process and identify potential challenges to adoption of model checks in VA.

\subsection{Analysis}
\label{sec:eval-analysis}
We characterized participants’ use of \system{} with and without model check functionality through a qualitative analysis of interview recordings and transcripts.
The interviewer reviewed each transcript and video, at first identifying episodes of interest that revealed patterns of visual data exploration behavior, uses of model checks, and potential improvements to \system{}.
The study team then analyzed these episodes of interest and summarized what participants said in terms of topical themes and design tensions, focusing especially on any difficulties that participants seemed to have expressing 
or interpreting 
\revision{model checks.}

\begin{figure}[t]
    \centering
    \includegraphics[width=3in]{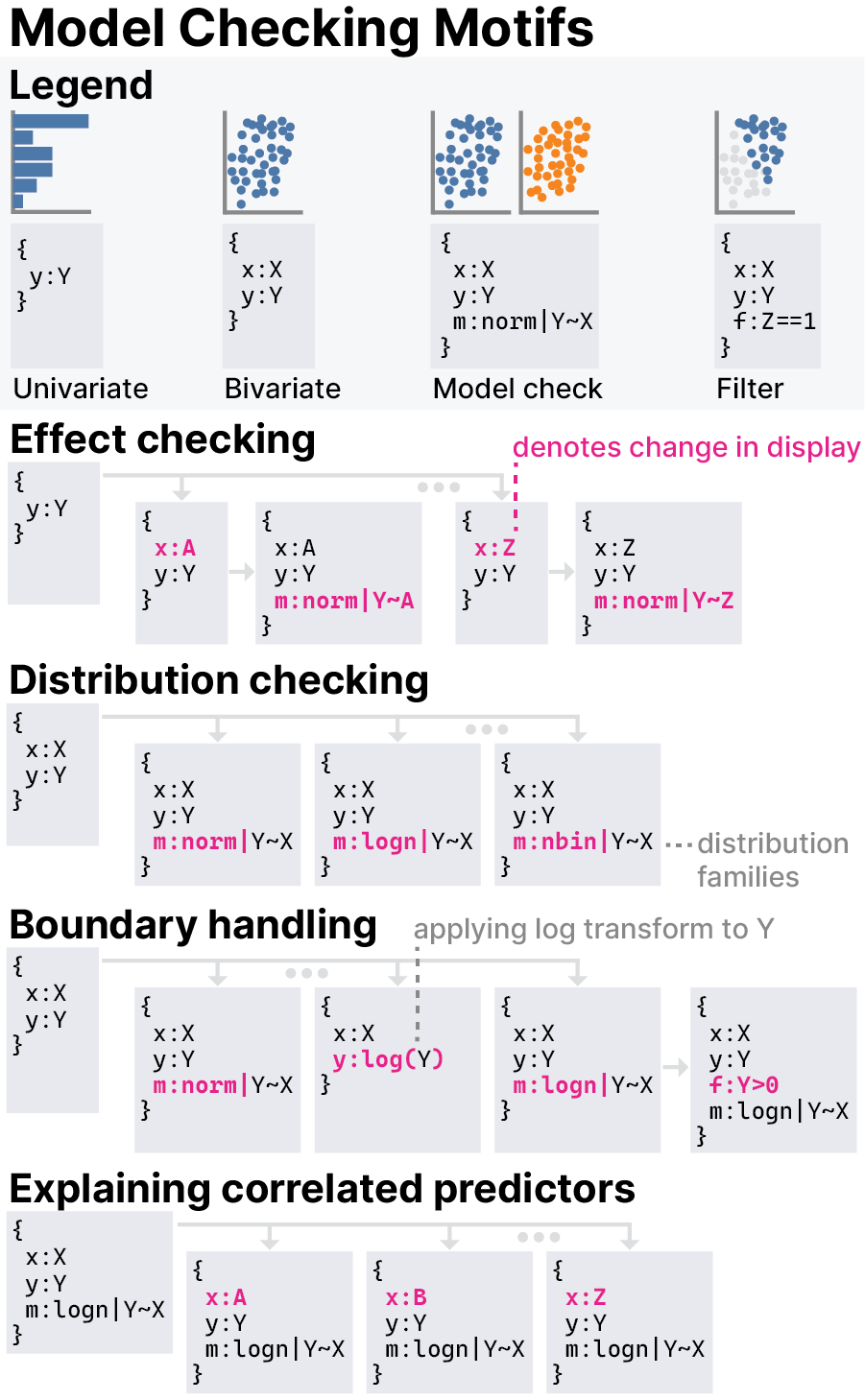}
    \setlength{\abovecaptionskip}{3pt}
    \setlength{\belowcaptionskip}{-14pt}
    \caption{
        Model checking motifs observed in our evaluation of \system{}.
    }
    \label{fig:motifs}
\end{figure}

\subsection{Results}
\label{sec:eval-results}
We report on how patterns of data exploration behavior in a prototype VA tool differ with versus without model checking functionality.
We also summarize what visual cues participants relied on to interpret model check visualizations and challenges that participants encountered using \system's interface, with an eye toward the development of future VA tools incorporating model checks.

\subsubsection{Data analysis motifs}
Our analysis revealed patterns of visual data exploration behavior that we refer to as \textit{``motifs''}, which were common sequences of operations in \system{} that often seemed to serve the purpose of making data exploration systematic.
For example, Figure~\ref{fig:motifs} shows motifs specific to model checking (described below).
Our notion of motifs is similar to Battle and Heer's ``VA subtasks''~\cite{battle2019-characterizing-eva}.
However, unlike subtasks, motifs were not focused on answering questions specific to a given dataset, but rather these were procedures we observed participants apply repeatedly to surface and check relationships between different subsets of variables. 
These motifs help us describe users' analysis behavior at baseline---when using a simple drag-and-drop VA tool without model checking functionality---and then to describe the new behaviors that emerge when using model checks as implemented in \system{}.

\noindent
\textbf{Behavior \textit{without} model checks.}
Common visual exploration motifs we observed without model checking enabled facilitate use cases such as anomaly detection and pattern finding.
Some of these motifs are similar to patterns of analysis behavior described in prior work.
For example, these include the \textbf{univariate tour} viewing univariate summaries of each variable (7/12 participants) and the \textbf{bivariate tour} making pairwise comparisons of all predictors to look for correlations (6/12 participants).
These are similar to automated summaries provided by previous systems like Voyager2~\cite{voyager2017}, which enable the user to get \textit{``a cross tab of all the covariates'' (P07)}.
Some participants (4/12) worked even more systematically, applying these motifs to selected \textit{``clusters of explanatory variables'' (P10)} in turn.

Other motifs we observed without model checking functionality focused on explaining the distribution of a specific outcome variable of interest.
For example, we observed \textbf{hunting for main effects} by cycling through each predictor variable comparing it to the outcome in a bivariate view (7/12 participants), which is sometimes interleaved with a univariate tour of predictors, and \textbf{hunting for interactions} by faceting bivariate plots of main effects by a sequence of third variables to look for interaction effects (9/12 participants).
Another motif used by most participants (9/12) to account for conditional structure in the data was \textbf{filter toggling}, eyeballing a pattern before and after applying variations on a particular filter, which was also described in prior work on view sequencing in narrative visualizations~\cite{hullman2013-sequences}. 
All participants referred to expectations at some point when performing these distribution-explaining motifs, and some participants (8/12) would tell stories about the data in order to explain discovered patterns.

\noindent
\textbf{Behavior \textit{with} model checks.}
When we introduced model checks to \system{} in the second data exploration session, we noticed marked changes in the behavior of most participants.
Without model checks enabled, all participants except \textit{P12} briefly explored patterns across a broad set of available variables and then circled back to recheck relationships they had investigated previously. 
However, with model checks enabled, sequences of related operations became longer, and data exploration became less circuitous,
\revision{consistent with findings of prior work that adding modeling functionality to VA tools leads to less breadth of analysis during insight-oriented data exploration~\cite{Koonchanok2021}.} 

Model checking tended to structure participants' thinking around one or two long chains of operations geared toward gradually improving models.
Some of these model improvement motifs were concerned with finding an appropriate way to approximate the distribution of the outcome variable.
For example (Fig.~\ref{fig:motifs}), we observed participants (7/12) \textbf{distribution checking} to hone in on a plausible distribution family and (5/12 participants) \textbf{boundary handling} by iteratively applying different distribution families, transforms, and filters in order to account for natural boundaries in the data (e.g., no counts below zero).
Other model improvement motifs were more concerned with predictor selection.
Similar to 
\revision{the demonstrated \textit{understand-correlated-predictors} use case (see Section~\ref{sec:system-scenario}),}
we observed some participants (6/12) \textbf{explaining correlated predictors} by checking the patterns predicted by a provisional model against the domains of predictors \textit{not} included in the model to see whether any structure in the data remains unaccounted for. 

Although many of our participants exhibited these model improvement motifs, such motifs did not seem to benefit all participants equally.
A subset of participants (5/12) used model improvement motifs to focus on developing a fine-grained understanding of the data generating process (DGP), demonstrating the kind of thinking we designed \system's model checking functionality to elicit in users.
For example, one participant said both that, \textit{``When I fit a model, I was definitely thinking more about the second moment.'' (P09)} and that, 
\begin{quote}
\textit{``With the initial [think-aloud session], I didn't think about bounds as much... I don't think it came up, but it was only when trying to describe a model that I started putting these theoretical bounds on what values could take.'' (P09)} 
\end{quote} 
In addition to paying increased attention to variance relationships and expected boundaries, these participants (5/12) mentioned that visual model checks calibrated their sense of uncertainty around outliers, undersampled regions of the data, and the tails of distributions. 
HOPs depicting uncertainty as sampled model predictions seem to clarify the possible structure of the data in regions where data is sparse.

However, for a smaller subset of participants, relying on the same model improvement motifs promoted \textit{fixation} on trying to understand the underlying implementation of model checks (2/12 participants) or on superficial matching of the shape of the predictive distribution to the shape of the data without thought about what the relationships meant (2/12 participants).
Two participants fixated on long sequences of model improvements and neglected to explore most of the dataset during the \textit{think-aloud with model checks} portion of the interview.

Exceptions to this trend of longer operation sequences 
were two participants who always used model checks in a manner more akin to one-off statistical tests for specific relationships.
They would specify models with versus without an 
\revision{effect of}
a particular predictor 
in order to see which model's predictive distribution seemed more in line with the data, similar to the 
\revision{demonstrated}
\textit{evaluate-the-plausibility-of-an-\revision{interpretation}} 
use case 
\revision{(see Section~\ref{sec:system-scenario}).}
We call this motif \textbf{effect checking} (Fig.~\ref{fig:motifs}).
Most participants (8/12) used
model check \revision{visualizations}
as provisional hypothesis tests at some point, interleaving visual analysis motifs for pattern discovery such as \textit{hunting for main effects/interactions} with model checking motifs for vetting 
\revision{data interpretations}
such as \textit{effect checking} or \textit{explaining correlated predictors}.



\begin{figure}[t]
    \centering
    \includegraphics[width=3in]{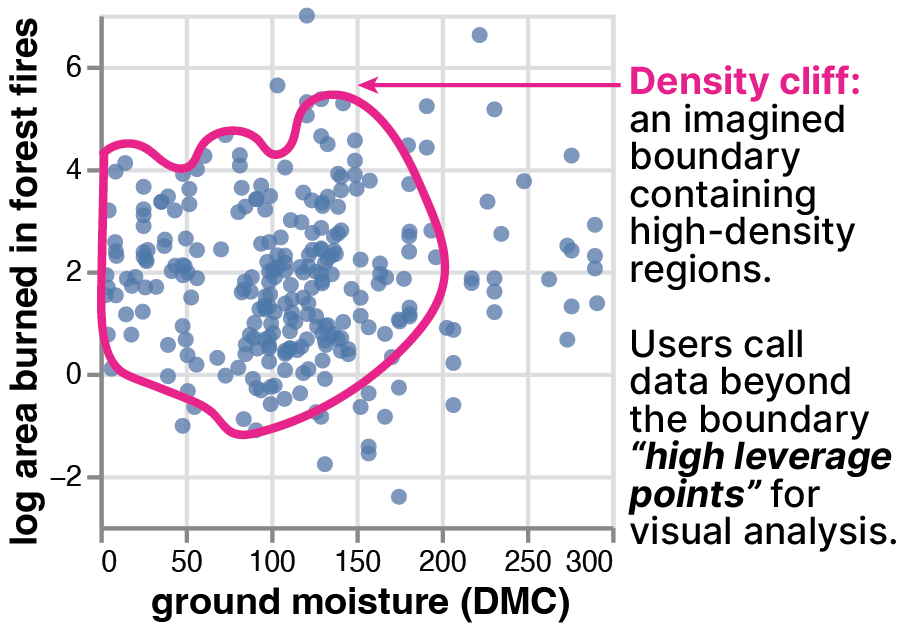}
    \setlength{\abovecaptionskip}{3pt}
    \setlength{\belowcaptionskip}{-14pt}
    \caption{Depiction of an imagined boundary we refer to as a ``density cliff''. Many heuristics for visual analysis with scatterplots and strip plots seem anchored to this kind of reference.}
    \label{fig:density-cliff}
\end{figure}

\begin{figure}[b]
    \vspace{-10pt}
    \centering
    \includegraphics[width=3in]{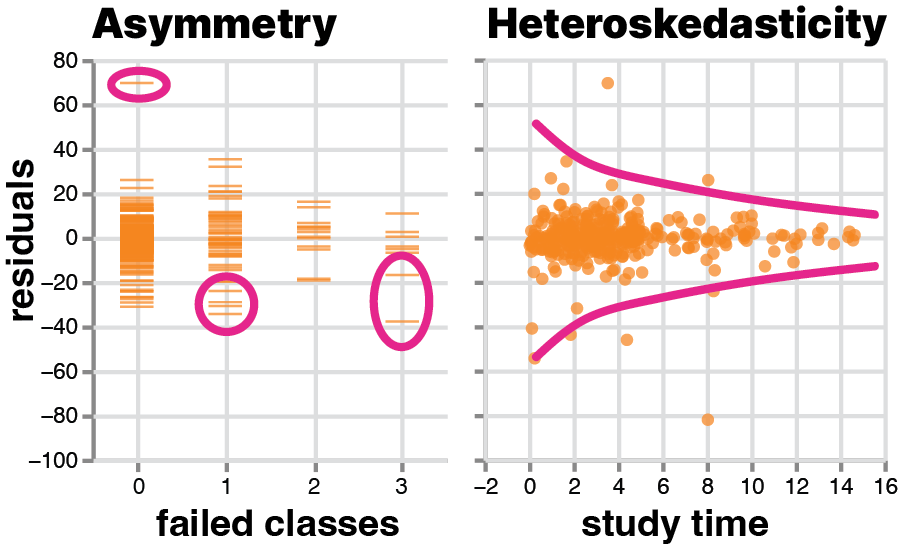}
    \setlength{\abovecaptionskip}{3pt}
    \setlength{\belowcaptionskip}{-14pt}
    \caption{Visual cues that participants used to interpret residuals.}
    \label{fig:resid-cues}
\end{figure}

\subsubsection{Interpretation of model check visualizations}
Most participants (8/12) interpreted model check visualizations primarily in terms of the match between the shape of the data and model predictions.
The most salient visual cues for shape seemed to involve the concentration of data points in \system's scatterplots and strip plots, with many participants (8/12) paying special attention to an imaginary boundary around high-concentration regions, which we refer to as a \textit{``density cliff''} (Fig.~\ref{fig:density-cliff}).
Some participants seemed self-aware about how points near a density cliff 
influenced
their perception of a pattern, for example, \textit{``I do see that again these kind of larger [fires] are occurring... but that's really cherry picking just a few high-leverage points, though, so I'm not sure that means a whole lot.'' (P01).}
When data points were further away from a density cliff, some participants (3/12) referred to them as separate modes and others (8/12) referred to them as outliers depending on the number of points 
and how far they were from the density cliff.
\revision{Although prior work finds that ``regression by eye'' down-weights outliers~\cite{Correll2017-regression-by-eye}, our participants judged misfit using heuristics that were more sensitive to anomalies.}
When the shapes did not match between the data and model predictions, participants (4/12) interpreted this as a sign of misfit.
A subset of participants frequently relied on residuals plots to assess model fit, with some (2/12 participants) describing asymmetry around zero as a helpful cue and others (4/12 participants) pointing out  heteroskedasticity of prediction errors across the domain of a given predictor as a cue that the relationship with that predictor was not adequately captured by the model (Fig.~\ref{fig:resid-cues}).

\subsubsection{Challenges using model checks in \system{}}
Many difficulties participants faced with \system{} involved distrust of eyeballing.
One participant bemoaned when struggling to visually differentiate the fit quality of two models, \textit{``My lying eyes sometimes deceive me.'' (P01).} 
This issue is certainly not unique to model checks---indeed, most VA workflows tend to rely on such subtle visual inferences.
However, all participants described similar dilemmas and wanted to supplement visual inferences with diagnostic metrics for models (e.g., information criteria, R-squared, regression coefficients).
We avoided providing such diagnostics in \system{}
\revision{in part}
because we worried that participants would over-rely on them in ways that would interfere with our design goal of promoting generative thinking (see Section~\ref{sec:design}) and detract from our investigation of visual model checking.

Similarly, most participants (9/12) wanted the ability to derive summary statistics such as counts, means, and quartiles on the fly and to apply them to regions of the chart using brush interactions.
For example, two participants wanted to brush out a region of a scatterplot and compute the number of data points in that region relative to the size of the dataset. 
Many (7/12 participants) wanted the ability to change the default visualizations of the tool by binning continuous variables or by recasting variables as different data types.
These difficulties seemed to stem from the 
lack of granularity that \system's scatterplots and strip plots provide for inspecting the relative density of regions on charts: \textit{``It's either dense, medium dense, or not very dense.'' (P03)}. 
Supporting histograms and density plots would have addressed some of these concerns, however, emphasizing visual aggregation may also lead to overconfidence in visual inferences~\cite{Nguyen2020}, so 
\revision{defaults must navigate this trade-off.}
Seeking stronger visual signals about relative density led two participants to rely on highly inefficient exploration strategies such as scrolling through many faceted bar charts. 

When using the model bar to express provisional 
models, most problems stem from the challenge of anticipating what 
account\revision{s} for misfit.
Although some participants (5/12) said that model checks in \system{} make it quick to try out models, and others (6/12 participants) said visual model checks made it easy to see misfit, they seemed to struggle to improve models using the interface.
For example, all participants began modeling with a normal distribution, even though the outcome variable in both datasets had a lower bound at zero, making it likely that the modeling assumption of Gaussian distributed residuals would be violated.
Upon seeing the resulting misfit, all participants except \textit{P11} at first added more variables to their model specification rather than changing their choice of distribution family.
Eventually, by using distribution checking or boundary handling motifs for iterative model checking, all participants except \textit{P12} discovered that the choice of distribution family accounted for misfit.
Participants may have added predictors before changing distributional assumptions because \system's model bar makes it easy to dump additional predictors into subsequent model iterations, and they did not stop to rethink distributional assumptions.

Many participants (7/12) struggled when choosing among distribution families,
\revision{which was often a contributor to misfit.}
This was especially common when trying to reason about the back-transformations (see Section~\ref{sec:system-implementation}) in the log normal and logit normal families.
Multiple participants attributed some difficulties to lack of knowledge about the domain of the datasets (3/12) or to being rusty at specifying regression models in R (5/12).
\revision{In contrast to prior work showing that VA users don't follow up on model misfit~\cite{Choi2019a}, we find that all participants attempted to reason about what might be wrong with misfit models.}

Our results suggest the need for future work on guided model elicitation interfaces.
Two participants suggested scaffolding a path of model exploration between the simplest possible intercept model and a ``dredge model'' including all available predictors.
This is in line with practices in a Bayesian workflow~\cite{gabry2019,gelman2020bayesian} as well as the desire of some participants (5/12) to avoid including multiple highly-correlated predictors in their model specifications, exemplified by this quote:
\begin{quote}
    \textit{``How do we get the most parsimonious model? How do we remove things that are perhaps highly correlated, or even collinear, and have the best model with the fewest features? which is I think where I would go next, being able to prune the model a bit, not have things that I don't need.'' (P08)}.
\end{quote}
A few participants (3/12) noted situations where a single model could not account for distinct sub-populations within the data and requested support for partitioning the data into subsets to be modeled separately.
\section{Discussion}
\label{sec:discussion}
Our work building and evaluating \system{} points to new design requirements for model checks beyond those identified in Section~\ref{sec:design}, 
as well as articulatory gaps faced by users of visual analysis tools more broadly.

We find that \textbf{model checking can improve understanding of data generating process (DGP), but only when users avoid fixating on non-conceptual aspects of analysis}, such as the underlying implementation of statistical models or superficially matching model predictions to patterns in the data. 
Participants who used model checks to discover important attributes of the DGP such as boundedness or correlated predictors tended to be experienced at interpreting the relationship between modeling assumptions and data.
In contrast, the smaller subset of participants who fixated on non-conceptual aspects of model checking were more familiar with use cases for statistical models that prioritize predictive accuracy over scrutinizing assumptions. 
This partially confirms our design hypothesis about the utility of model checks in visual analytics, but it also points to the need for model checking tools to accommodate users with fluency in different modeling approaches.

Although model checks help users identify misfit in models, \textbf{users require guidance about which incremental improvements to a model could plausibly improve fit}.
For example, most participants added new predictors to a given model before considering whether they had chosen an appropriate distribution family for the data.
Future work might generate modeling recommendations that nudge users to \revision{(re-)}examine specific assumptions
\revision{or asserted structures}. 
More expressive tools that allow users to articulate specific aspects of misfit that concern them as input to a recommender are also well motivated---e.g., if a cluster of data points appear in the data distribution but not in model predictions.


Additionally,
we present \textbf{analysis \textit{``motifs''} that reflect procedural abstractions for visual analytics (VA) workflows}. 
Similar to patterns of visualization sequencing described in prior work~\cite{battle2019-characterizing-eva,hullman2013-sequences} and automated data summaries provided by previous systems such as Voyager2~\cite{voyager2017}, we think of these motifs as a 
\revision{workflow}-level
abstraction describing subroutines in visual analysis.
In contrast, 
\revision{a chart}-level abstraction
such as the grammar of graphics~\cite{Wilkinson2005} 
focus\revision{es} on 
specifying analyses 
\revision{one visualization at a time.}
For our purposes, these motifs serve to characterize common sequences of operations within VA sessions.
%
Looking to the future of VA tools, we envision interfaces that enable users to author and reuse motifs in data exploration workflows,
\revision{extending the idea of ``wildcard fields'' in Voyager2~\cite{voyager2017} such that users could specify a sequence of diagnostic visualizations and apply them across a series of provisional models reflecting competing data interpretations.}
Specifying model checking workflows at the level of motifs could further reduce the amount of effort required to rigorously check 
\revision{data interpretations}
without loss of expressiveness.

Importantly, beyond need-finding for model checking interfaces, we also 
\revision{add to a line of work~\cite{Choi2019a,Choi2019b,Koonchanok2021} demonstrating}
that \textbf{model checks can integrate cleanly into exploratory visual analysis workflows}, offering support to users transitioning between open-ended exploration tasks~\cite{battle2019-characterizing-eva} and what Tukey~\cite{tukey1972data} calls ``rough confirmatory analysis.''
Participants who spend relatively little time working with programming interfaces appreciated that \system{} offered access to modeling and diagnostics approaches which usually require scripting in statistical programming languages.
For them, \system{} offered a quick way to refine 
\revision{data interpretations}
and fight \textit{``a false sense of security about the quality of the data'' (P02)} that they say sets in when using graphical user interfaces for exploratory visual analysis. 
Similarly, participants who spend relatively more time using statistical tools viewed \system{} as a way to test their preconceived notions by rapidly iterating on models.
One of these participants described how, \textit{``[Visual model checking] felt much less p-hacky than it might have if I'd been looking at the numbers. You know, I'm not just choosing the [model] with the best R-squared or whatever.'' (P08)}, echoing others for whom the ease of generating visual checks enabled a faster pace of analysis without loss of rigor.

\subsection{Ongoing \& future work}
\label{sec:discussion-future}
Investigating how to integrate \textbf{model recommendations} into a tool like \system{} is a natural follow-up to our work.
Recommendations should account for what is known about the user's understanding of the DGP at the time of recommendation, similar to the way that tools like Tisane~\cite{Jun2022tisane} and Visual Causality Analyst~\cite{Wang2016,Wang2018} anchor model suggestions on knowledge elicited from users. 
If the user's preferred model at a given moment during analysis represents their traversal of a \textit{``model space''} they are searching, recommendations should be proximal to and informed by the user's current model.
\revision{Rather than suggesting a single ``best'' proximal model for the user to consider next, recommender systems should highlight multiplicity, where multiple alternative models perform similarly and cannot be easily distinguished.}
Model recommendations should also be informed by the visual model checks an analyst creates and the specific patterns in data that they struggle to explain.
We envision interfaces where analysts can directly select a pattern in a visualization that they wish to better explain---i.e., the kind of superficial pattern matching that some participants fixated on---and a recommendation engine would suggest additional 
model\revision{s}
that could capture the pattern.
This could help analysts remain focused on how visual patterns relate to their conceptual understanding of DGP, providing softer on-ramps to the kind of generative thinking that \system{} facilitates.

\begin{figure}[t]
    \centering
    \includegraphics[width=3in]{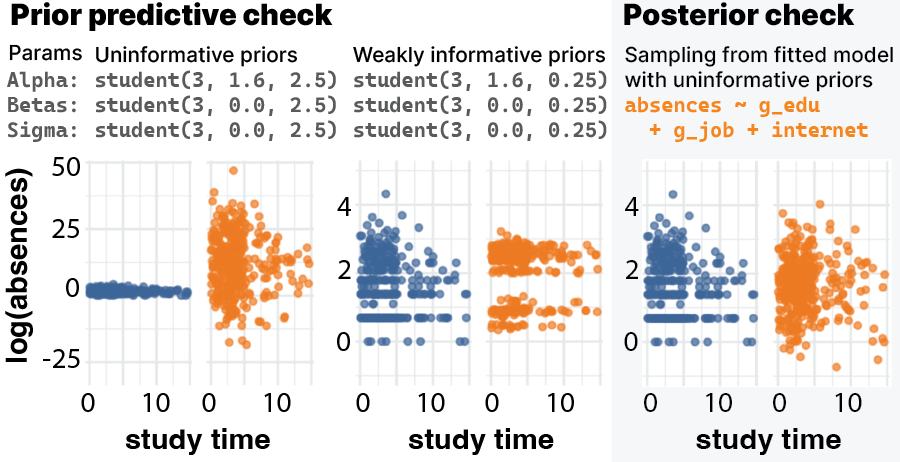}
    \setlength{\abovecaptionskip}{3pt}
    \setlength{\belowcaptionskip}{-14pt}
    \caption{Examples of prior vs posterior predictive checks for the same model. This model check examines whether study time is predictive of log absences after accounting for proxies of socioeconomic status. Uninformative priors (left) blow up the scale of the y-axis, making an unhelpful model check. Sampling from weakly informative priors (middle), rather than a fitted model (right), can lead to model checks that show a larger space of possible patterns that follow from a model's assumptions (e.g., homogeneity of variance), potentially increasing visual signal for judging whether a model is mis-specified.}
    \label{fig:prior-vs-post-mc}
\end{figure}

In \system{}, we implemented comparisons between observed data and predictions from a \textit{fitted} model. 
However, there are many possible \textbf{ways to sample predictions} reflecting a given 
\revision{model,}
and the specific approach plays a large role in determining the visual appearance of a model check.
\revision{Specifying a model defines} 
a \textit{parameter space}, a latent multivariate distribution that is sampled from and further transformed in order to produce a prediction.
Sampling fitted models as we did in \system{} means sampling from this distribution in ways that minimize the discrepancy between the data and model predictions.
However, Bayesian prior predictive checking offers an alternative approach, which helps to shed light on the importance of these low-level implementation choices (Fig.~\ref{fig:prior-vs-post-mc}, see Supplemental Materials). 
In a prior predictive check, the choice of prior determines the scale of the predictive distribution, enabling this approach to show more possible structures in predictions that 
\revision{are consistent with a model specification}
but which are ruled out by the fitting process.
Contrasting these approaches raises the question of whether enabling users to specify priors and sample from them would be a more direct way to assess 
\revision{users' expectations}
through visual checks,
\revision{to the extent that elicited models represent users' beliefs as assumed by prior work~\cite{Choi2019a,Choi2019b,Koonchanok2021}.} 
However, designing for Bayesian predictive checks might increase the level of modeling complexity beyond what some VA users are familiar with, e.g., requiring procedures for fine tuning and justifying priors.
Where concerns about model check implementation details threaten to pull analysts’ attention away from DGP, we suggest providing more documentation of how a tool like \system{} samples predictive distributions in order to provide transparency.

Critical to these endeavors is incorporation of \textbf{diagnostic metrics} for ensuring that a seemingly well-fitting model also has predictive power, such as cross validation on a holdout set. 
Using holdout sets rather than training data for model checks would increase the visual discrepancy between data and model predictions, providing stronger visual signals of potential misfit.
Having access to diagnostics like cross validation metrics might help address the lack of confidence in eyeballing among some study participants, in addition to helping users of future tools like \system{} avoid 
\revision{overfit explanations,}
overconfidence in mis-specified models, and 
\revision{misleading local extrema in model search space.}

\revision{
Related to issues of overfitting and iterative model development, it is crucial to develop \textbf{safeguards against post-hoc statistical inference} in tools like \system{}. 
A potential failure mode with any VA tool is that users will make multiple comparisons before deciding what relationships are important~\cite{Zgraggen2018}, which can inflate false positive rates similar to p-hacking~\cite{Simmons2011}.
Anticipating the concern that incorporating modeling into VA tools could exacerbate these risks, we carefully design \system{} to make it difficult to perform confirmatory hypothesis testing---e.g., by not providing p-values, learned coefficients, or other metrics conventional in statistical testing.
Unlike previous work, which enforces a strict order of operations where VA tools only elicit models before showing the data~\cite{Choi2019a,Choi2019b,Koonchanok2021}, we opt for a more open-ended study of how people use modeling in VA when left to their own discretion.
In the user study, we find that participants are relatively cautious about overtrusting models they fit, suggesting they do not view model check visualizations as a definitive inference method.
However, we observe some participants using effect checking motifs (Fig.~\ref{fig:motifs}) in ways that resemble hypothesis testing, and this suggests opportunities for design patterns that distinguish between comparisons in service of model development versus proper confirmatory testing.
As proposed above and in prior work~\cite{Zgraggen2018}, future tools like \system{} should support holdout sets for model validation and confirmatory testing, which will require careful consideration of how to avoid leaking information when swapping training and holdout sets.
In cases where holdout sets are unavailable, we envision that users could save models to test on future data when it becomes available.
Future work should evaluate various regimes for controlling post-hoc inference, including algorithmic approaches (e.g.,~\cite{Berk2013,Tibshirani2016}), to determine which are best suited to the VA setting.
}

Additionally, future research should develop tools for ensuring effective 
\revision{model check visualizations.}
Systems like \system{}, along with prior work~\cite{hullman2021hdsr,mcnutt2022grammar}, motivate a \textbf{model check grammar} to facilitate greater flexibility in traversing the design space of 
\revision{uncertainty visualizations required for model checking.}
This would be necessary in order to enable motif-based authoring systems where analysts create visual checks and apply them across a sequence of related views demarcated by, e.g., different variable selections, data transformations, or model iterations.
How a visual check should change when model predictions are grafted onto it is determined by constraints that are difficult to express using Vega~\cite{vega2016} due to the diffuseness and viscosity of its notation~\cite{green1990cognitive-dimensions}.
As a workaround, when implementing \system{} we created a separate Vega template for each possible layout, similar to recent work on ``parameterized declarative templates''~\cite{mcnutt2021-ivy}.
However, extending this approach for model checks requires additional research and development on abstractions for 
\revision{model check visualizations}
in particular.
To support engineering future tools, a model check grammar should include (1) layout constraints defining how data and model predictions 
\revision{are}
treated, (2) primitives for 
sampling from models in order to generate uncertainty visualizations and model diagnostics, and (3) interaction techniques that enable users to express which patterns reflect conceptually important misfit. 
More broadly, understanding what makes a model check
\revision{visualization}
effective may not be equivalent to what makes a visualization of observed data effective, motivating empirical work.

\subsection{Limitations}
\label{sec:discussion-limitations}
Large scale and high-dimensional data are open problems in VA tools that remain unaddressed by \system{}.
For the purpose of creating a proof-of-concept tool, we focused on model checks for relatively small datasets.
However, some of our design choices 
\revision{need refinement to work}
at larger scales.
For example, 
\revision{using HOPs~\cite{Hullman2015} for higher-dimensional views requires careful interaction design (e.g., in node-link diagrams~\cite{Zhang2022-NetHOPs}).}
\revision{Relatedly,}
disaggregated views~\cite{Nguyen2020} showing one mark per data point become unwieldy at large sample sizes, due to limitations 
\revision{around}
visual crowding~\cite{Alvarez2011}
and computer memory.
Aggregation is a common approach to side-stepping these issues of scale, however, aggregated data and model outputs have fundamentally different meanings as summary statistics than disaggregated data and model predictions.
Future work will need to resolve when \revision{such aggregation} 
is and is not advisable.

Similarly, our choice to limit \system{} to position encodings (i.e., x-axis, y-axis, row, column) rules out visualization techniques that might be more suitable for higher-dimensional data, where datasets have many variables that can each take on many values.
\system{} only enables viewing a subset of a high-dimensional dataset's features at one time, a limitation of many VA tools.
Although 
\revision{model check visualizations}
carry information about variables that are not currently in view---i.e., a model can make predictions based on a variable that isn't visualized---analysts may still struggle to reason about the complex structure of correlated variables that underlies a particular view.
Future work 
should more directly investigate whether using model checks to reason about variables that are not in view can solve the high-dimensionality problem, and the ways in which this approach might be error prone.


Some conditions of our user study design were difficult to control. 
When preparing datasets for participants to explore, we dropped certain variables and observations from the student absences dataset in order to make it match the size of the forest fires dataset.
It's possible that these adjustments affected participants' ability to find patterns in the data, however, this did not come up explicitly during our interviews.
Beyond the size of datasets, there were other factors which likely impact our results that were not possible to control for.
These include each participant's level of interest in or familiarity with the provided datasets.

\section{Conclusion}
\label{sec:conclusion}
We present \system{}, a proof-of-concept tool enabling analysts to express and check statistical models during visual data exploration.
\system{} is a design investigation into how 
\revision{visual analytics (VA)}
tools can incorporate statistical modeling to facilitate more rigorous thinking about a data generating process (DGP).
This augments the typical process of visual pattern discovery with procedures for articulating and scrutinizing claims about a DGP, which we argue elevates 
\revision{VA}
tools from producing nebulous ``insights'' to vetting 
\revision{provisional data interpretations}
and providing a more concrete basis for further analysis.
Our work demonstrates the potential of 
\revision{model check visualizations}
to better connect 
\revision{VA tools}
with the cognitive and statistical procedures by which analysts develop and evaluate their conceptual understanding of data.

\acknowledgments{%
  We thank Justin Talbot and Andrew Gelman for providing early input about design requirements for incorporating regression models into graphical user interfaces for exploratory visual analysis. Jessica Hullman thanks NSF \#2211939 and \#1930642 for supporting this work.
}

\bibliographystyle{abbrv-doi-hyperref}

\bibliography{evm-ref}


\appendix 


\end{document}